\documentclass[aps, prl, showpacs, superscriptaddress, ctexart, nofootinbib, twocolumn]{revtex4}
\usepackage{amssymb,amsmath,graphicx,color,microtype}

\usepackage{color}

\begin{document}
\title{Testing quantum gravity effects with latest CMB observations}


\author{Yi-Fu Cai}
\email{yifucai@physics.mcgill.ca}
\affiliation{Department of Physics, McGill University, Montr\'eal, QC, H3A 2T8, Canada}

\author{Yi Wang}
\email{yw366@cam.ac.uk}
\affiliation{Centre for Theoretical Cosmology, DAMTP, University of Cambridge, Cambridge CB3 0WA, UK}

\begin{abstract}
Inspired by quantum gravitational physics, the approach of non-commutative (NC) phase space leads to a modified dispersion relation of gravitational waves. This feature, if applied to the very early universe, gives rise to a modified power spectrum of primordial tensor perturbations with a suppression of power on large scales. We confront this phenomenon with the BICEP2 and Planck experiments, and show that inflation with the modified dispersion relation can simultaneously fit the observations better than the standard inflationary paradigm. In particular, the numerical result implies that with the latest cosmological microwave background (CMB) observations, a quantum gravity modified power spectrum of primordial tensor modes is preferred at a statistical significance of more than $3\sigma$ compared with the minimal model. Our study indicates that the potential tension between the BICEP2 and Planck data may be resolved by quantum gravity effects.
\end{abstract}

\pacs{98.80.Es, 98.80.Cq}

\maketitle


{\it Introduction.---}
Recently, the BICEP2 collaboration reported a detection of primordial B-mode polarization on the CMB at more than $5\sigma$ confidence level (C.L.)~\cite{Ade:2014xna}. This measurement indicates that, if all the B-mode signal is from primordial gravitational waves, the inflationary tensor-to-scalar ratio is constrained as $ r = 0.2 ^{+0.07}_{-0.05} ~ (68\%~{\rm C.L.}) $. This result is in tension with the Planck data reported in 2013 (Planck13 hereafter), i.e., $r <0.11 ~ (95\%~{\rm C.L.})$~\cite{Ade:2013zuv}. Thus, the standard inflationary paradigm is severely challenged in simultaneously explaining these two observations.

Theoretically, it has been learned that $r$ measures the variation of the inflaton field during inflation through the relation $\Delta\phi/\Delta N = \sqrt{2r} M_p /4$ \cite{Lyth:1996im}. The BICEP2 result then implies that inflation requires a super-Planck field variation and hence all Planck suppressed non-renormalizable operators could become important already during inflation. In this regard, it is natural to examine implications of quantum gravity phenomenology in the very early universe, in particular the effects of non-commutativity \cite{Snyder:1946qz}, as inspired by fundamental theories such as string theory~\cite{Seiberg:1999vs, Li:1996rp, Connes:1997cr} and the Generalized Uncertainty Principle~\cite{Amati:1988tn, Maggiore:1993rv}. Moreover, a so-called noncommutative field theory, which describes an effective action that contains noncommutative terms partly obtained from noncommutative geometry, was developed extensively in the literature \cite{Carroll:2001ws, Carmona:2002iv, Gamboa:2005bf, Falomir:2005it}.

In this Letter, we examine observational signatures of a quantum gravity inspired NC effect on the power spectrum of primordial gravitational waves which are relevant to the current CMB observations. Specifically, we propose a modified parametrization for the power spectrum of primordial tensor modes based on the formalism of phase space non-commutativity. The key feature of this spectrum is that its amplitude is conserved and nearly scale-invariant at small scales but obtains a dramatic suppression at large scales~\cite{Cai:2007xr}. Making use of the CosmoMC package~\cite{Lewis:2002ah}, we confront the theoretical parametrization with the WMAP9, Planck13 and BICEP2 data. Our result reveals that the NC dispersion relation yields a better fit to the combined CMB observations compared with standard predictions. Therefore, our study hints at new physics beyond the classical General Relativistic inflationary paradigm of Big Bang cosmology. Also it provides an observational window to probe quantum gravity effects in present cosmological observations.


{\it Theory setup.---}
We begin with a brief discussion of primordial gravitational waves assuming NC phase space in the framework of a flat FRW universe. This effect is known to be associated with non-commutative space and its corresponding quantum field theory description requires at least two internal degrees of freedom, for instance, a complex scalar~\cite{Carmona:2002iv} or the graviton~\cite{Cai:2007xr}. With that, in the context of single field inflationary cosmology, it does not affect the scalar sector but only changes the tensor modes. In particular, it appears in the quantization process of tensor modes and leads to a modified dispersion relation which gives rise to a feature in the primordial tensor spectrum: the spectral index of tensor modes changes from scale invariant at small scales to blue at large scales. Thus, the amplitude of primordial tensor fluctuations exiting the inflationary Hubble radius during the first several e-folds decreases.

In the flat FRW background, the tensor fluctuations are given by $\bar h_{ij}$ by expanding the metric as follows,
\begin{eqnarray}
ds^2=a(\tau)^2[-d\tau^2+(\delta_{ij}+\bar h_{ij})dx^idx^j]~,
\end{eqnarray}
where $\tau$ is the comoving time, $a(\tau)$ is the scale factor, and the Latin indexes represent spatial coordinates. For simplicity, we redefine the fluctuations through $h_{ij} = a \bar h_{ij}$.
Using comoving time, the linearized Lagrangian of the cosmic tensor fluctuations is expressed as,
\begin{eqnarray}
 {\cal L}= \frac{1}{4}[h_{ij}'^2+\frac{a''}{a}h_{ij}^2-(\partial_l h_{ij})^2] ~.
\end{eqnarray}
The tensor fluctuations $h_{ij}$ satisfy the relations $h_{ij}=h_{ji}$ and $h_{ii}=h_{ij,j}=0$. By defining the conjugate momenta $ \pi_{ij} \equiv {\delta}{\cal L}/{\delta h_{ij}'} = \frac{1}{2}h_{ij}' $, one can canonically quantize the system using the standard equal-time commutation relation: $[h_{ij}(\tau, {\bf k}), \pi_{kl}(\tau, {\bf k}')] = \frac{1}{2}(\delta_{ik}\delta_{jl}+\delta_{il}\delta_{jk}) \delta^{(3)}({\bf k}-{\bf k'})$. However, in order to include the non-commutativity into the above system, it is natural to make a deformation on the commutation relation of the conjugate momenta as: $[\pi_{ij}(\tau, {\bf k}), \pi_{kl}(\tau, {\bf k}')] = \alpha_{ijkl} \delta^{(3)}({\bf k}-{\bf k'})$, where $\alpha_{ijkl}$ is a constant matrix. In order to be consistent with the algebraic properties of tensor fluctuations, $\alpha_{ijkl}$ satisfies $\alpha_{ijkl} =\alpha_{jikl} =-\alpha_{klij}$ and thus yields $\alpha_{ijkl} =\alpha_{ik} \delta_{jl} +\alpha_{il}\delta_{jk} +\alpha_{jk}\delta_{il} +\alpha_{jl}\delta_{ik}$ with $\alpha_{jk} =-\epsilon_{0jkl} \alpha^l$, where $\epsilon_{ijkl}$ is a totally antisymmetric tensor and $\alpha^l$ a constant 3-vector.

Note that the deformed equation of motion for primordial tensor modes in Fourier space is given by $ h_{ij}'' +k^2h_{ij} -\frac{a''}{a}h_{ij} +8\alpha_m\epsilon^{0lim}h_{jl}' = 0 $. For convenience, we take $\alpha_m =\{0,0,\alpha_3\}$ and introduce two independent tensor modes as: $v_1=\frac{1}{\sqrt{2}}(h_{11}+ih_{12})$, $v_2=\frac{1}{\sqrt{2}}(h_{11}-ih_{12})$. Then, their equations of motion are given by,
\begin{eqnarray}
\label{v1}
 v_j''+(-)^{j-1} 8i\alpha_3v_j'+k^2 v_j-\frac{a''}{a}v_j=0~,
\end{eqnarray}
with $j=\{1,2\}$, respectively. Note that the case $\alpha_3=0$ reduces to commutative space without quantum gravity effect where the second term of eq. \eqref{v1} vanishes, and then the regular dispersion relation is recovered.
By imposing vacuum initial condition, the above equations can be explicitly solved to yield:
\begin{align}\label{exact v1}
 v_j(k, \tau) = \frac{\sqrt{\pi}}{2} e^{i(\nu+\frac{1}{2}) \frac{\pi}{2}} (-\tau)^{\frac{1}{2}} e^{(-)^j4i\alpha_3\tau} H_{\nu}^{(1)}(-l\tau)~,
\end{align}
in terms of the $\nu$-th Hankel function of the first kind with $\nu = \frac{\epsilon -3}{2(\epsilon -1)}$, and $\epsilon \equiv -\frac{\dot H}{H^2}$ is the slow roll parameter. We introduce an effective frequency of the tensor modes to be $l\equiv \sqrt{k^2+16\alpha_3^2}$.

Following the definition of the power spectrum $P_T(k) \equiv \frac{2k^3}{\pi^2M_p^2} \sum_{j=1,2}|\frac{v_j}{a}|^2$, one can substitute the above solutions and obtain the tensor spectrum on super-Hubble scales. For convenience of relating the theoretical prediction with observations, we propose the following template for data analysis,
\begin{align} \label{eq:paraSpectrum}
  P_T(k) = A_T \left( \frac{k}{k_\mathrm{pivot}} \right)^{n_t} \frac{ k^{3}}{(k^2 + \alpha^2 k^2_\mathrm{pivot})^{3/2}}~,
\end{align}
where $A_T \equiv 2H^2/(\pi^2M_p^2) $ is defined as the amplitude of the tensor spectrum at a pivot scale, \footnote{Consider that $n_t$ is suppressed by the slow roll parameter, the choice of the pivot scale mainly affects the analysis of the factor $(k^2 + \alpha^2 k^2_\mathrm{pivot})^{3/2}$ and therefore the redefinition of $\alpha$.} and $n_t$ is the corresponding spectral index. The term $\alpha^2 k^2_\mathrm{pivot}$ appearing in the denominator is the signal arising from quantum gravity effects. In the case of NC phase space, the parameter $\alpha$ is associated with the coefficient $\alpha_3$ via $\alpha^2 k^2_\mathrm{pivot} = 16\alpha_3^2$, and we name it as the NC parameter. From the above expression, one easily observes that the tensor spectrum is nearly scale-invariant in the large $k$ regime. This is in agreement with the result obtained in standard slow roll inflation. However, for the tensor modes with $k\ll \alpha k_\mathrm{pivot}$, the spectrum becomes very blue and hence, its amplitude decreases on large scales. Moreover, when $\alpha=0$, the spectrum reduces to the regular result in standard inflationary paradigm. As usual, one can introduce a tensor-to-scalar ratio $r\equiv A_T/A_S$, with $A_S$ being the amplitude of the power spectrum for curvature perturbations. Since primordial curvature perturbations reply on the inflaton fluctuations which are in the scalar sector, $A_S$ is the same as the regular one derived in the standard inflationary paradigm.

\begin{figure}[htbp]
\centering
\includegraphics[width=0.4\textwidth]{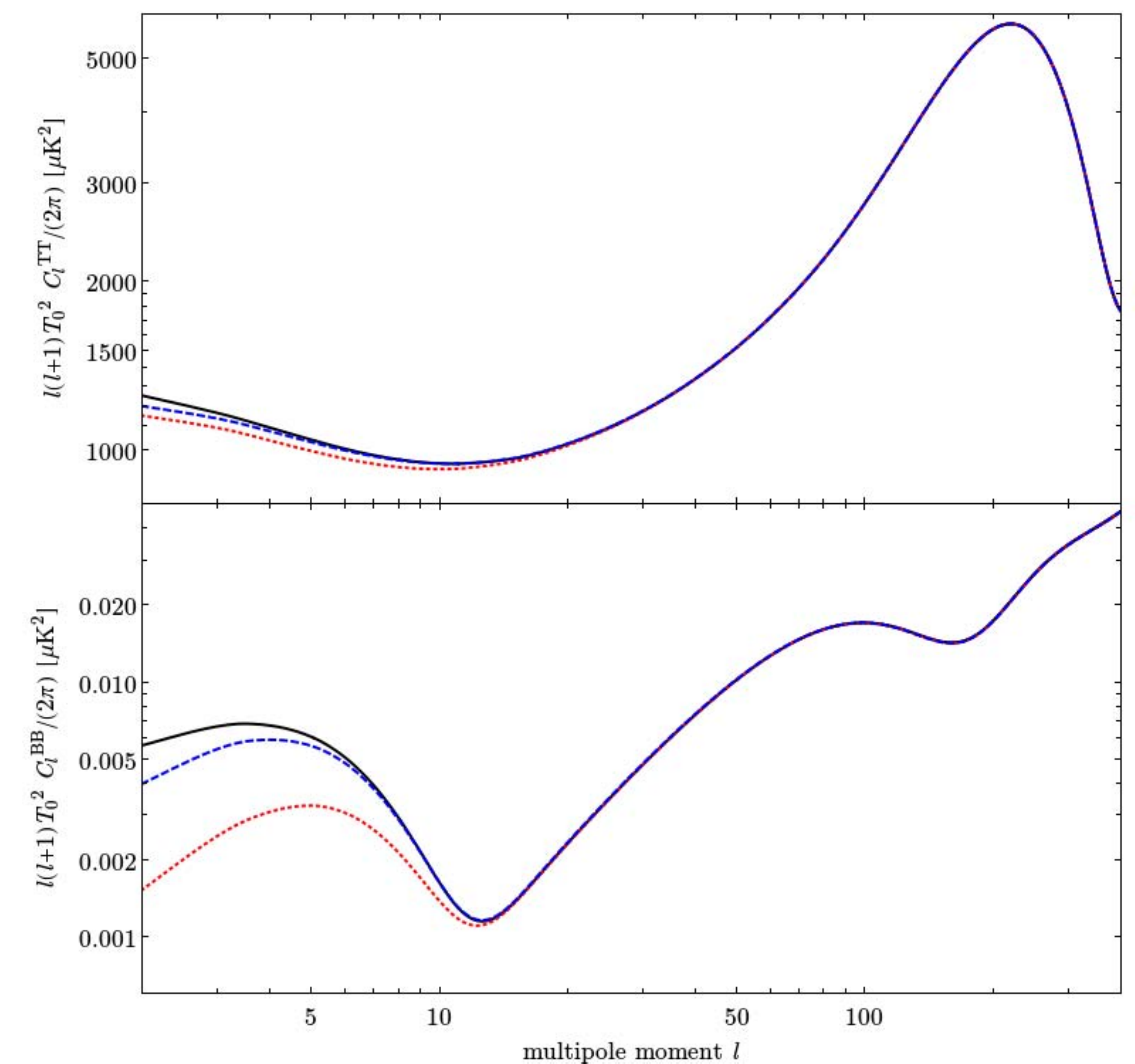}
\caption{\label{fig:comp} Signatures of the quantum gravity effect on the primordial TT (upper panel) and BB (lower panel) power spectra with different values of $\alpha$. We compare a $\Lambda$CDM model (black solid line) with two NC modified inflation models with different values of the NC parameter as described in the main text. }
\end{figure}

{\it Numerics.---}
To test the plausible quantum gravity effect in CMB experiments, we take $k_\mathrm{pivot} = 0.01$Mpc$^{-1}$ so that it accommodates the BICEP2 observation. Fig.~\ref{fig:comp} illustrates this impact on the CMB temperature and tensor spectra with different values of the NC parameter. We consider two NC modified inflation models with $\alpha=0.02$ (the blue dashed line) and $\alpha=0.05$ (red dotted line), respectively. At large angular scales, the power spectra are affected by the quantum gravity effect, so the decrease on the amplitudes happens when $k<\alpha k_\mathrm{pivot}$. Moreover, the bigger the value of $\alpha$ we choose, the more prominent is the signature of this suppression at low $\ell$ values.

It is known that the BICEP2 data strongly favors non-vanishing primordial tensor perturbations, namely $ r = 0.2 ^{+0.07}_{-0.05}$ in the intermediate $\ell$ regime ($50 < \ell < 100$). This nonzero value of $r$, if it still exists in the low $\ell$ regime, can lead to extra power in the CMB temperature spectrum and make it difficult to fit to the low $\ell$ Planck data. As a result, there exists a tension between these two data sets. In the NC modified model, however, without changing the spectrum of primordial curvature perturbations, we have shown in the lower panel of Fig.~\ref{fig:comp} that the amplitude of the tensor spectrum is suppressed due to the quantum gravity effect. Correspondingly, while one can generate enough large primordial B mode to fit to the BICEP2 data, the amplitude of the same spectrum damps in the low $\ell$ regime and, therefore, can very well explain the Planck data simultaneously.

Given the above theoretical setup, we perform a Markov Chain Monte Carlo (MCMC) fitting using a generalized CosmoMC package, which has been modified to calculate the CMB power spectra under the quantum gravity inspired NC effect. The cosmological parameters are: $(\Omega_{b}h^{2}, \Omega_{m}h^{2}, \Theta_s, \tau, n_s, A_S, r, n_t, \alpha) $, in which $\Omega_{b}$ and $\Omega_{m}$ are the baryon and cold dark matter densities relative to the critical density, $\Theta_{s}$ is the ratio (multiplied by 100) of the sound horizon to the angular diameter distance at decoupling, $\tau$ is the optical depth to re-ionization, $n_s$ is the spectral index and $A_S$ is the amplitude of curvature perturbation at the pivot scale, $r$ is identical to the tensor-to-scalar ratio at small scales as explained previously, and $n_t$ is the corresponding spectral index of tensor modes. Moreover, we have one additional parameter $\alpha$ which is associated with the quantum gravity effect in the NC modified model.

\begin{figure}[htbp]
\centering
\includegraphics[width=0.155\textwidth]{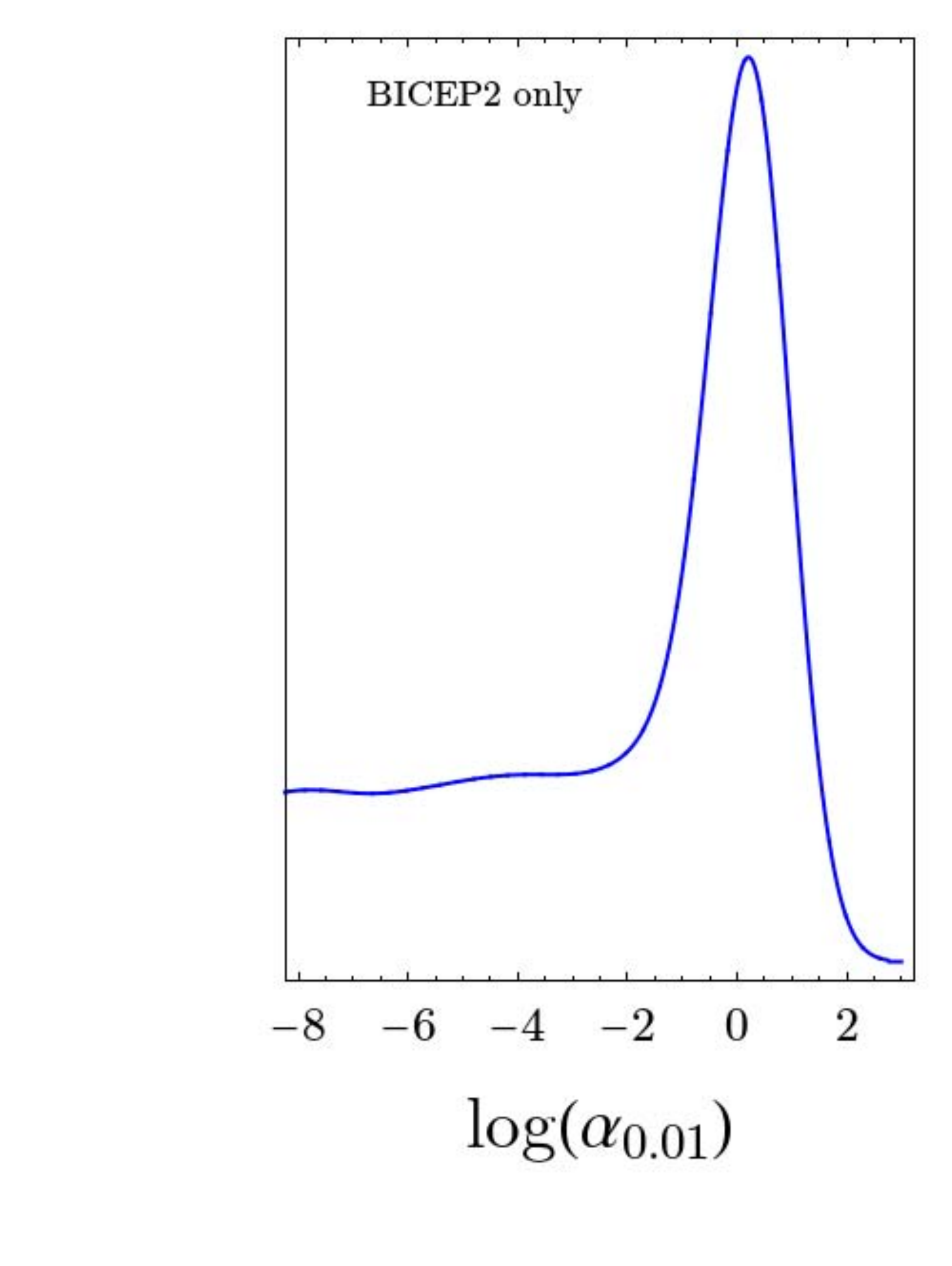}
\includegraphics[width=0.155\textwidth]{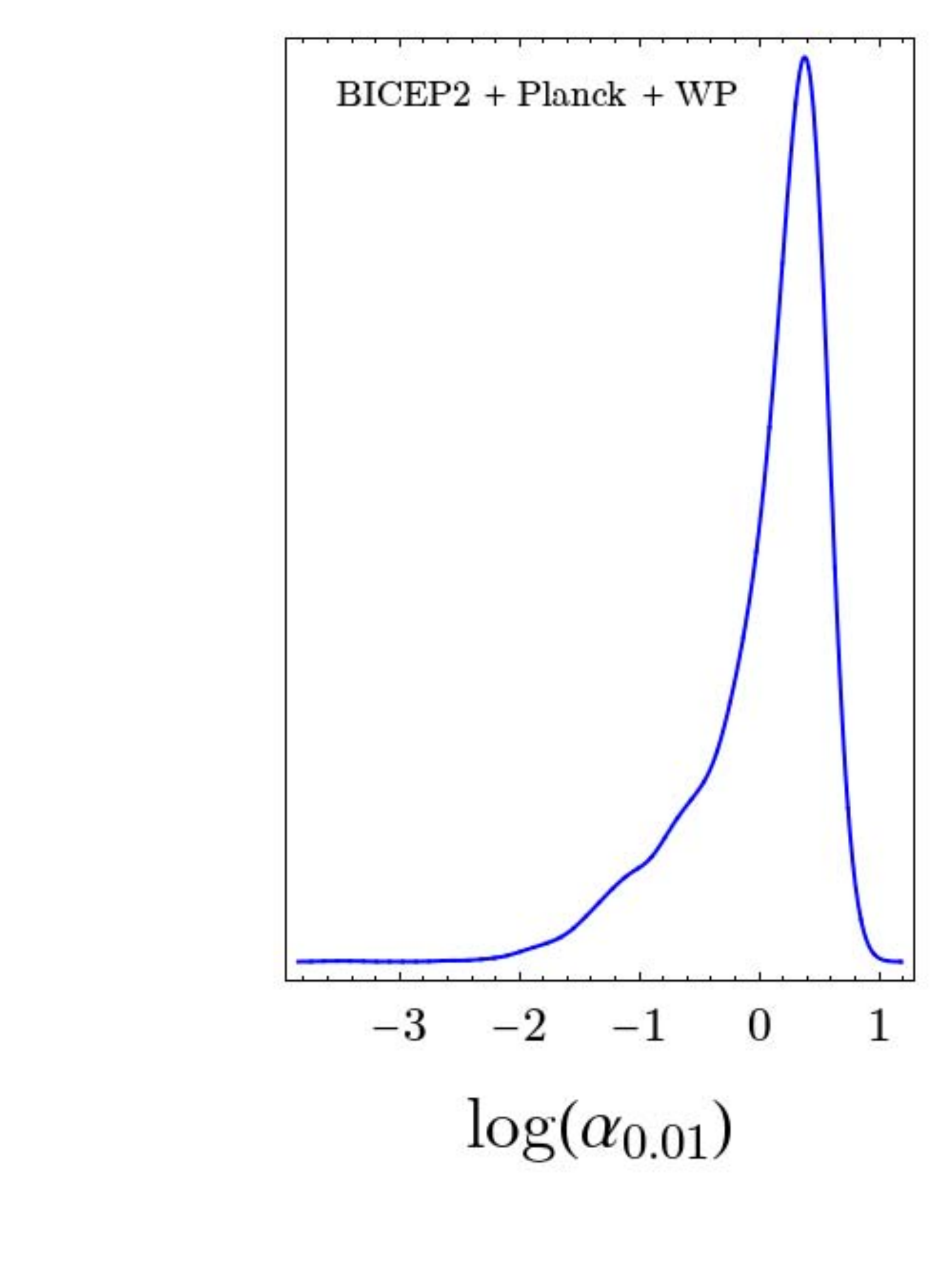}
\includegraphics[width=0.155\textwidth]{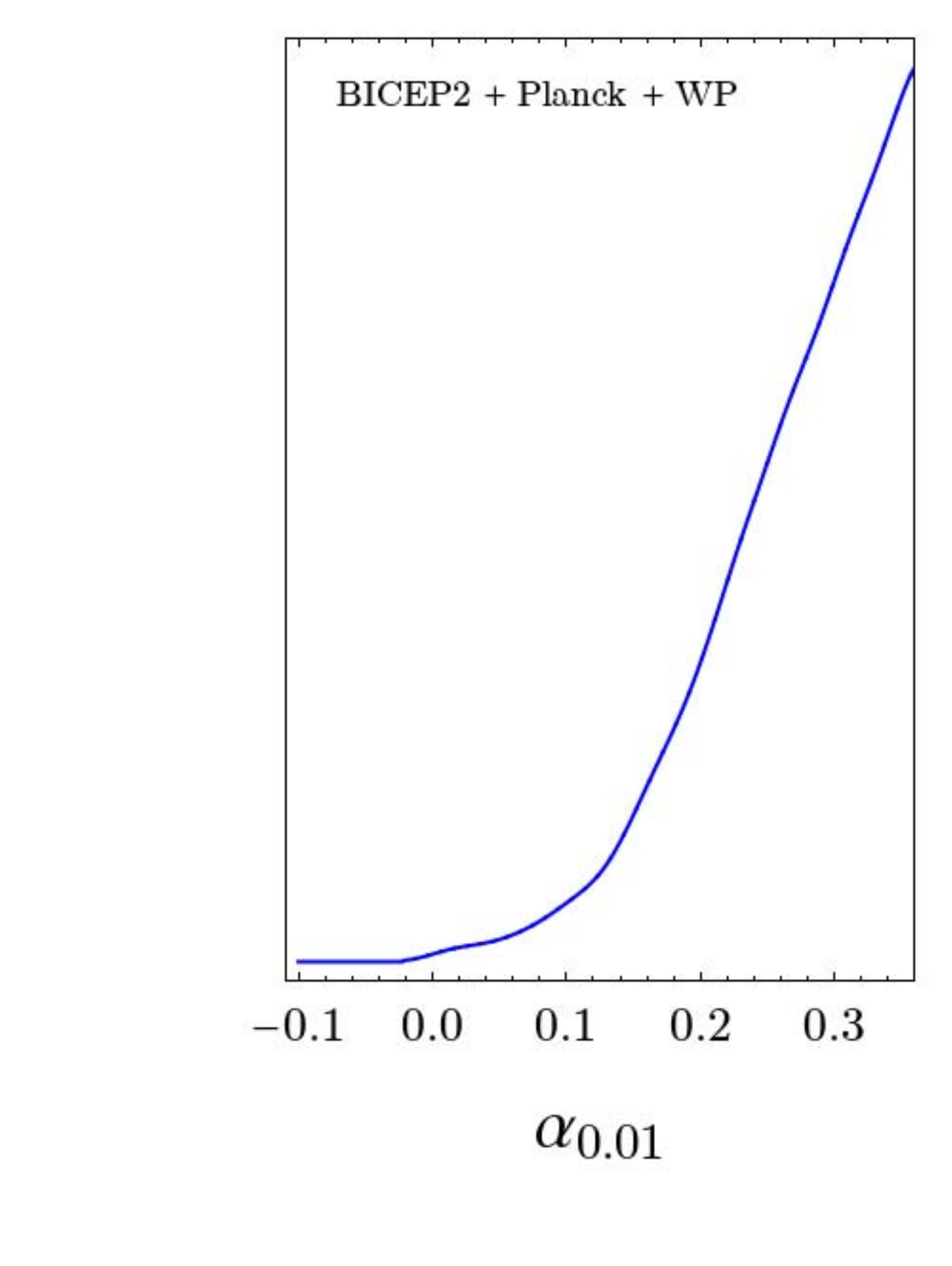}
\includegraphics[width=0.155\textwidth]{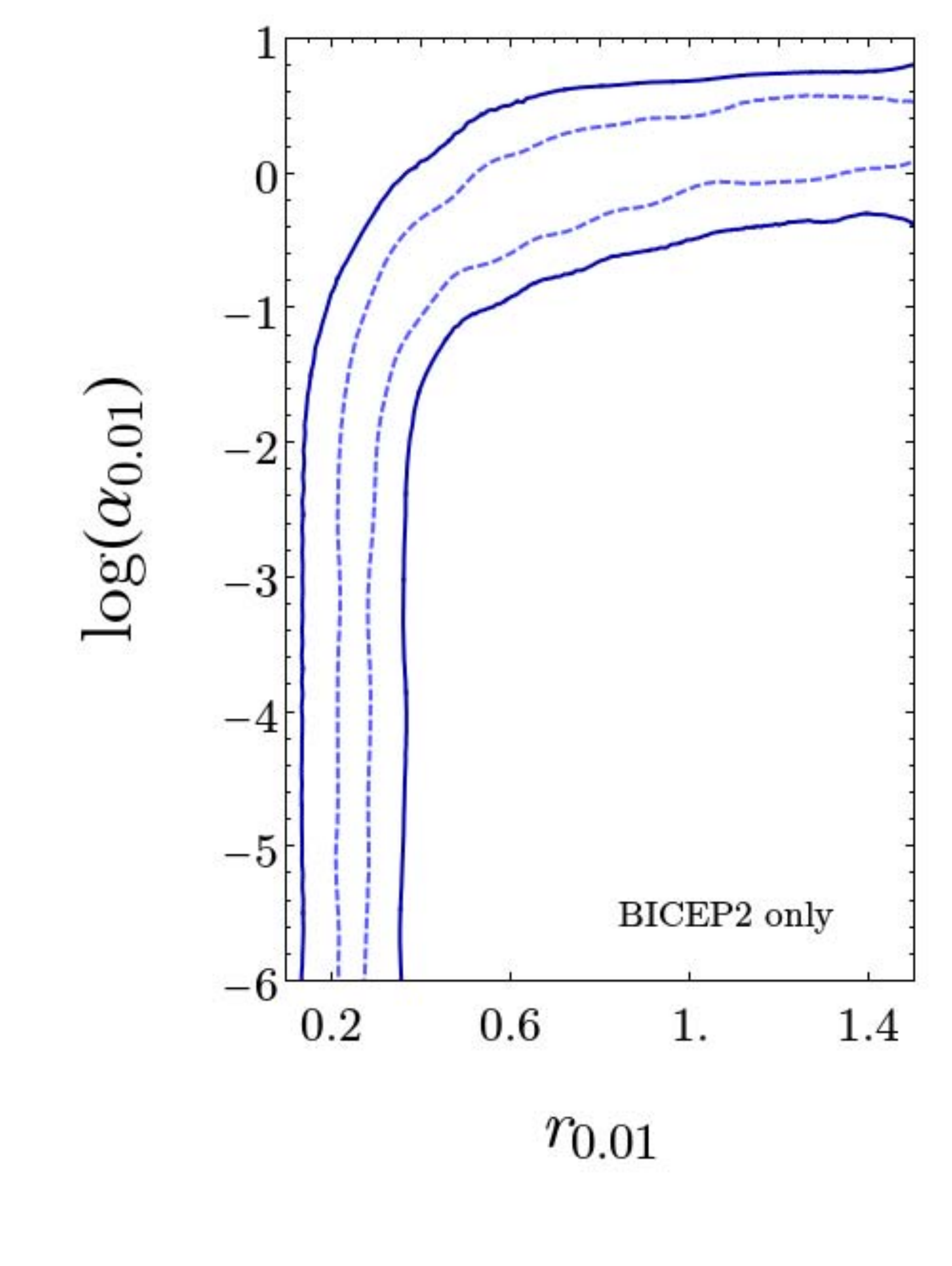}
\includegraphics[width=0.155\textwidth]{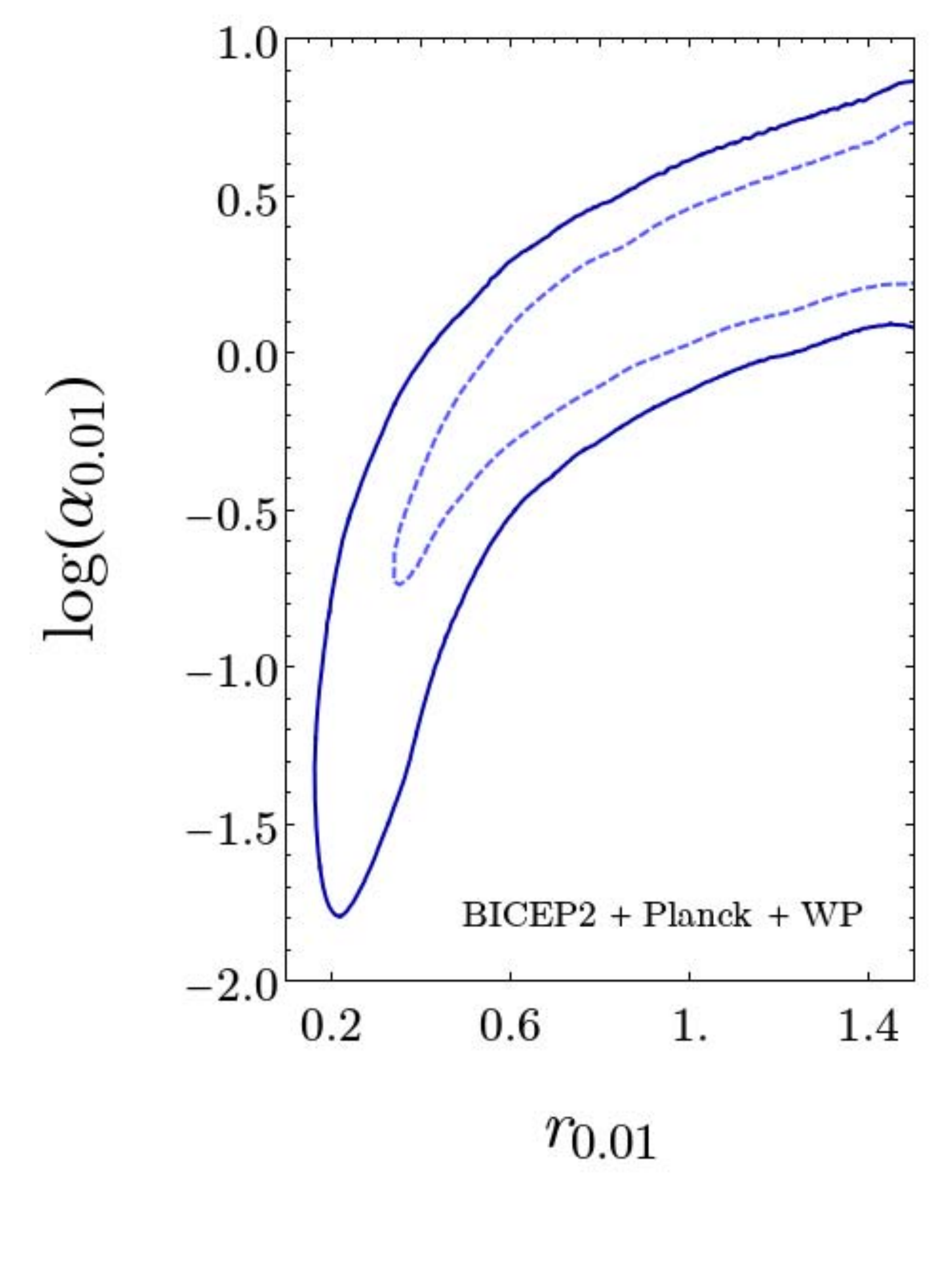}
\includegraphics[width=0.155\textwidth]{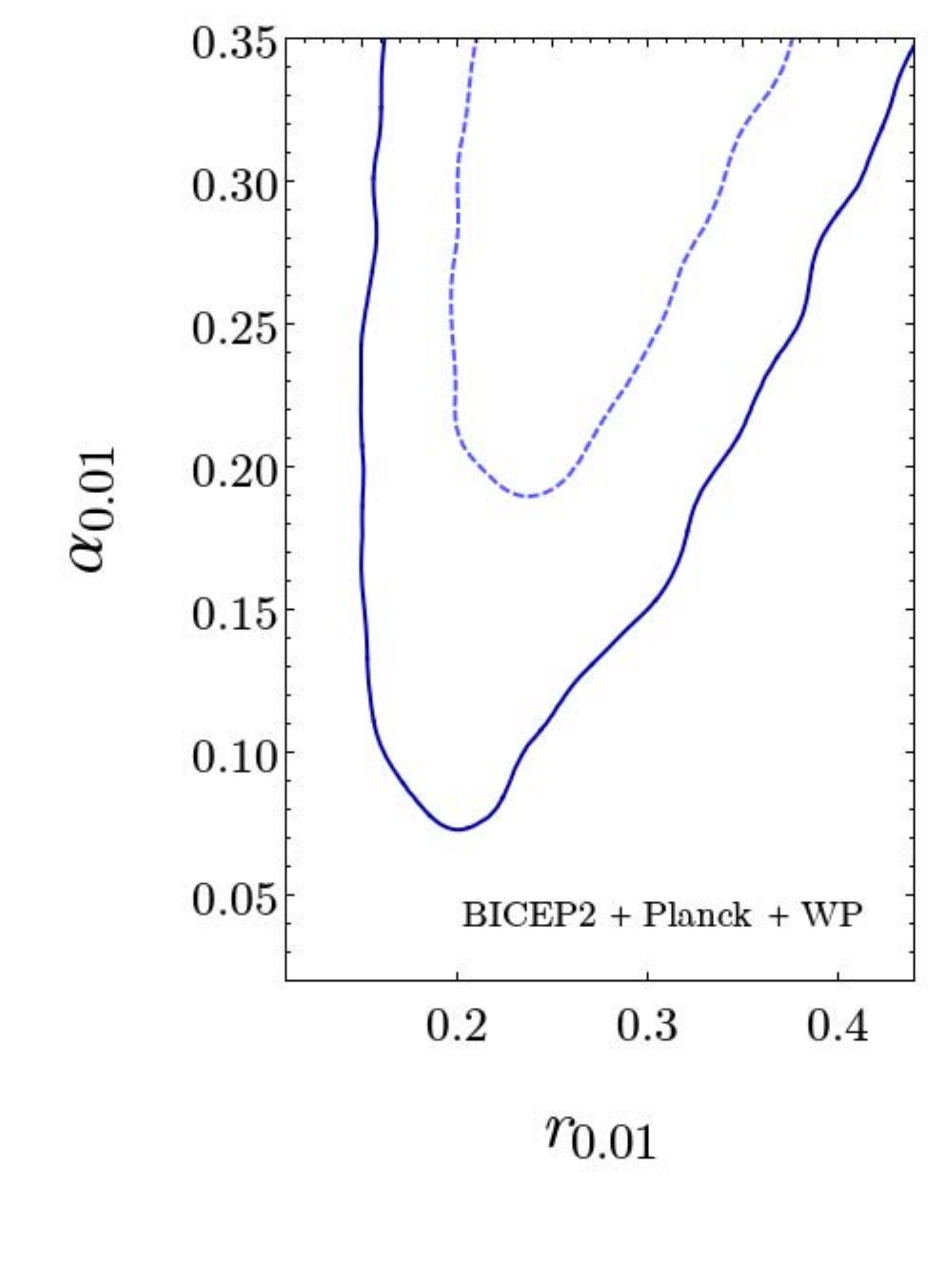}
\caption{\label{fig:contour} One and two dimensional constraints on the NC parameter $\alpha$ and the tensor-to-scalar ratio $r$ at the pivot scale in the NC modified inflation model. }
\end{figure}

Particularly, we analyze the observational constraints on the model from three points of view in order to examine the consistency of the numerical results. First, we only use the BICEP2 data to constrain the correlation between the log function of the NC parameter $(\log\alpha)$ and the tensor-to-scalar ratio $r$; then, we combine the BICEP2 and Planck13+WMAP9 data to further constrain the correlation between $\log\alpha$ and $r$; and afterwards, we use the BICEP2+Planck13+WMAP9 data to perform the constraint on this correlation directly. Our numerical results are presented in Fig.~\ref{fig:contour}.

In the left column of Fig.~\ref{fig:contour}, we apply the BICEP2 data only and find that the $(r, \log\alpha)$ contour is not bounded from below and thus the standard inflation model without any quantum gravity corrections is consistent. However, after marginalizing over $r$ a value with $\alpha = 1.0$ is preferred, as seen from the peak of the probability distribution curve.

The constraint of this contour becomes significantly improved after including the Planck13+WMAP9 data as is shown in the lower panel of the middle column in Fig.~\ref{fig:contour}. The shape of the probability distribution curve is almost unchanged but the constraint is improved, i.e., $\alpha = 1.0 ^{+0.74}_{-0.43}$ at $1\sigma$. This result indicates a non-zero amplitude of the NC parameter which thus hints at new physics beyond the standard paradigm. The contour favors a positive correlation between $\log\alpha$ and $r$. This implies that, when we increase $\log\alpha$, the value of $r$ also becomes larger in order to compensate the suppression effect brought by the NC physics.

Note that in the first two columns of Fig.~\ref{fig:contour}, we have taken the log function of the NC parameter in the fitting since its parameter space can be dramatically compressed. However, there is the risk of involving a certain prior in the analysis using this method, i.e., the limit of $\alpha\rightarrow 0$ (corresponding to $\log\alpha\rightarrow -\infty$) is not accessible to the MCMC. Therefore, one ought to examine whether the previous results are reliable by fitting $\alpha$ in linear function directly. To check this issue, we also run the MCMC with the combined BICEP2+Planck13+WMAP9 by assuming $\alpha \in [0, 0.4]$. We find the best fit value of the NC parameter $\alpha$ approaches the prior boundary $\alpha = 0.4$ in this case and which is consistent with the previous results.

Along with constraining the favored regime of $\alpha$, it is also helpful to analyze how unlikely $\alpha<0.02$ would be, which corresponds to the case that the NC effect is unobservably small. Applying the chain rule of probability, one gets the probability $P(\alpha<0.02) = P(\alpha<0.02|\alpha<\alpha_0) ~ P(\alpha<\alpha_0)$ where $\alpha_0 = 0.35$ is chosen as an intermediate regime between the linear sampling and logarithm sampling method. Moreover, the analysis of the linear sampling gives $P(\alpha<0.02|\alpha<\alpha_0)=0.0028$ and the logarithm sampling yields $P(\alpha<\alpha_0)=0.068$. To insert them into the chain rule, we obtain $P(\alpha<0.02) = 2\times10^{-4}$, which indicates $\alpha<0.02$ is disfavored at $99.98\%$, or $3.7\sigma$ confidence level.

Eventually, our result shows that the latest CMB data prefers a quantum gravity modified power spectrum of primordial gravitational waves at more than $3\sigma$ confidence level. One should note that such a statistical significance disfavors the minimal inflation model considered by the BICEP2 and Planck teams. We notice that there indeed exist other models which fit the data equally well according to the likelihood statistics, for example if the tensor sector has a constant blue $n_t$ \cite{Gerbino:2014eqa, Wang:2014kqa, Smith:2014kka}. To further distinguish between models, future data is needed. On the other hand, it is also important to note that different foreground modeling may weaken the BICEP2-Planck tension, and as a consequence would lower the statistical significance of a modified dispersion relation.

The parameter $\alpha$ newly introduced in the modified tensor spectrum is positively correlated with $r$ and thus can further enlarge the parameter space of the tensor-to-scalar ratio. This feature can be applied to distinguish our model from others, namely, in the model of relaxing the spectral index of tensor modes one finds that $n_t$ is negatively correlated with $r$ \cite{cosmocoffee}. Moreover, the suppression feature that appeared in the low $\ell$ regime of the power spectrum of primordial B-mode is scale dependent and therefore is another critical signature which is distinguishable from other mechanisms, such as the matter-bounce inflation model~\cite{Xia:2014tda}, the non-Bunch-Davis inflation model~\cite{Ashoorioon:2014nta}, or the string gas cosmology~\cite{Brandenberger:2006xi}. In particular, the bounce inflation model generally modifies the power spectra of both tensor and scalar types \cite{Piao:2003zm} and therefore can be easily distinguished from our scenario through the measurement of the CMB temperature spectrum.


{\it Conclusion.---}
The recent detection of non-vanishing primordial B-mode reported by the BICEP2 collaboration leads to extra power in the CMB temperature spectrum within the standard inflationary paradigm. The $r=0.2$ best fit value indicates a super-Planck inflaton field variation and hence open the possibility that quantum gravity effects may already be relevant during inflation. Consider that the BICEP2 detection would involve the contribution from dust, the primordial tensor mode may have a smaller amplitude. For example, if $r=0.1$, from the tension between BICEP2 and Planck, it can be estimated that the statistical significance of non-vanishing $\alpha$ would drop by $1\sigma \sim 2\sigma$. We hope to perform a detailed study of this case once a realistic foreground model is available for likelihood analysis.

Inspired by quantum gravity theories, we examine a generalized parametrization for the power spectrum of primordial gravitational waves with an modified dispersion relation which may arise from phase space NC. We discover a scale dependent suppression of the TT and BB power spectra. This suppression can alleviate the tension between the BICEP2 and Planck observations. Comparing our numerical analysis with the latest CMB data, such a quantum gravity modified power spectrum of primordial tensor modes can interpret the combined observations better than the minimal model without modification of tensor dispersion relation.

We would like to end by highlighting the significance of examining the observable signals of the possible quantum gravity effect within the inflationary paradigm. In the present Letter we proposed to probe this effect via the CMB polarization measurements. It is interesting to notice that, the existence of quantum gravity effect could effectively modifies the dispersion relation of primordial perturbations and hence might leave observational signatures by generating large non-Gaussianities \cite{Ashoorioon:2011eg}. Additionally, primordial gravitational waves under the NC effect may yield the CMB statistical anisotropy with a quadrupole asymmetry \cite{Shiraishi:2014owa, Wang:2012fi} and hence could help to examine this effect via the forthcoming data. With the accumulated high precision cosmological observations, it seems promising to experimentally probe the fundamental physics effects that only occur at extremely UV scales by virtue of these highly improved techniques.


{\it Acknowledgments.---}
We thank R. Brandenberger, E. Ferreira and S.-C. Su for helpful discussions. YFC is supported in part by NSERC and by Physics Department at McGill. YW is supported by a Starting Grant of the European Research Council (ERC STG grant 279617), and the Stephen Hawking Advanced Fellowship. This work was undertaken on the COSMOS Shared Memory system at DAMTP, University of Cambridge operated on behalf of the STFC DiRAC HPC Facility. This equipment is funded by BIS National E-infrastructure capital grant ST/J005673/1 and STFC grants ST/H008586/1, ST/K00333X/1.

\end{document}